# Experimental Observation of Higher-Order Topological Anderson Insulators

Weixuan Zhang[1, 2*], Deyuan Zou[1*], Qingsong Pei[1], Wenjing He[2], Jiacheng Bao[2], Houjun Sun[2$], and Xiangdong Zhang[1+]

[1]Key Laboratory of advanced optoelectronic quantum architecture and measurements of Ministry of Education, Beijing Key Laboratory of Nanophotonics & Ultrafine Optoelectronic Systems, School of Physics, Beijing Institute of Technology, 100081, Beijing, China

[2] Beijing Key Laboratory of Millimeter wave and Terahertz Techniques, School of Information and Electronics, Beijing Institute of Technology, Beijing 100081, China

*These authors contributed equally to this work. [+$]Author to whom any correspondence should be addressed. E-mail: zhangxd@bit.edu.cn; sunhoujun@bit.edu.cn

## Abstract

Recently, a new family of symmetry-protected higher-order topological insulators has been proposed and was shown to host lower-dimensional boundary states. However, with the existence of the strong disorder in the bulk, the crystal symmetry is broken, and the associated corner states are disappeared. It is well known that the emergence of robust edge states and quantized transport can be induced by adding sufficient disorders into a topologically trivial insulator, that is the so-called topological Anderson insulator. The question is whether disorders can also cause the higher-order topological phase. This is not known so far, because interactions between disorders and the higher-order topological phases are completely different from those with the first-order topological system. Here, we demonstrate theoretically that the disorder-induced higher-order topological corner state and quantized fraction corner charge can appear in a modified Haldane model. In experiments, we construct the classical analog of such higher-order topological Anderson insulators using electric circuits and observe the disorder-induced corner state through the voltage measurement. Our work defies the conventional view that the disorder is detrimental to the higher-order topological phase, and offers a feasible platform to investigate the interaction between disorders and higher-order topological phases.

Topological physics has become one of the most fascinating reach areas in recent years [1-7]. Based on the bulk-boundary correspondence principle, the conventional topological phase is always featured by the boundary states with one-dimensional lower than the bulk that hosts them. Recently, a novel class of the symmetry-protected higher-order topological insulator that possesses lower-dimensional boundary states have been proposed [8-12] and realized in many systems, including solid materials [13, 14], mechanics [15], acoustics [16, 17], microwaves [18], photonics [19-21], and electrical circuits [22, 23]. The key property of both conventional and higher-order topological insulators is their robustness of boundary states against weak disorders in the bulk. However, under the strong disorder, the non-trivial topological-bandgap gets closed, and the system goes into the trivial phase associated with the Anderson localization [24].

Contrary to this conventional wisdom, the recent theoretical result indicates that the emergence of protected edge states and quantized transport can be induced, rather than inhibited, by adding sufficiently strong disorder to the bulk of a topologically trivial insulator [25]. This disorder-induced topological phase, called the topological Anderson insulator, has attracted intensive theoretical attention [26-30] and has been experimentally demonstrated in one-dimensional (1D) synthetic atomic wires [31], 1D sonic crystals [32], and 2D helical waveguides [33]. So far, the study of topological Anderson insulators is all focused on the first-order topological phase, while, the in-depth investigation on the disorder-induced higher-order topological insulators remains lacking. A direct idea is extending the method used for the first-order topological Anderson insulator to the higher-order cases. However, compared to the first-order topological phase, the functionality of the disorder is drastically altered when it interacts with the higher-order topological system. For example, it has been pointed out that the on-site disorder, which can transform the trivial system to the first-order topological insulator, is detrimental to the higher-order topology [34]. Hence, the current method for realizing the first-order topological Anderson insulators cannot be directly used to fulfill the higher-order topological Anderson insulator (HOTAI). Thus, it is straightforward to ask whether the HOTAI exists, and how to realize it in experiments?

Here, we provide a solution to realize the HOTAI. Our scheme is based on a modified Haldane model with different values of the intra-cell and inter-cell couplings. It is found that the topological phase transition from the anomalous quantum Hall phase to the second-order topological insulator appears when the sufficiently strong disorder of geometric phase for next-nearest-neighbor couplings is added into the bulk. Furthermore, the classical analog of the proposed HOTAI is realized experimentally using the electric circuit. The disorder-induced 0D corner state in the 2D circuit, which is verified through the circuit simulation and voltage measurement, manifests the appearance of HOTAIs.

We start with considering the modified Haldane model with different values of intra- and inter-cell couplings, as depicted in Fig. 1a. It is found that there are six sites in the unit cell, and four types of hopping named nearest-neighbor (NN) intra-cell coupling ($\gamma_1$, black solid lines), NN inter-cell coupling ($\gamma_2$, red solid lines), next-nearest-neighbor (NNN) intra-cell coupling ($\lambda_1 e^{\pm i\varphi}$, green dash lines), and NNN inter-cell coupling ($\lambda_2 e^{\pm i\varphi}$, yellow dash lines), respectively. This lattice model can be effectively described by a tight-binding Hamiltonian, which can be expressed as:

$$H=\sum_{i}Ua_i^{\dagger}a_i + \sum_{<i,j>}\gamma_{ij}a_i^{\dagger}a_j + \sum_{<<i,j>>}\lambda_{ij}e^{i\varphi}a_i^{\dagger}a_j + \text{h.c.} \quad (1)$$

with $a_i^{\dagger}$ ($a_i$) being the creation (annihilation) operator at the site $i$. $U$ is the onsite potential. $<\ldots>$ ($<<\ldots>>$) indicates that the summation is restricted within NN (NNN) sites. $\varphi$ is the geometrical phase of NNN couplings. Additionally, we have $\gamma_{i,j}=\gamma_1$ ($\gamma_{i,j}=\gamma_2$) and $\lambda_{i,j}=\lambda_1$ ($\lambda_{i,j}=\lambda_2$), when sites $i$ and $j$ belong to the same unit (adjacent units).

At first, we investigate the system without disorders, where parameters are set as $U=0$, $\gamma_1=1$, $\gamma_2=5$, $\lambda_1=0.4$, $\lambda_2=1.5$, and $\varphi=2\pi/3$. Figures 1b and 1c show the calculated eigen-spectrum of the open lattice, which contains a total of nineteen units with $C_6$ symmetry, and the distribution of associated zero-energy mode, respectively. It is seen that the gapless edge state exists, ensuring the system is topologically equivalent to the anomalous quantum Hall phase (AQHP) of the Haldane model [35]. Then, we introduce NNN coupling disorders, where the geometrical phase of NNN coupling at different lattice sites is set as $\varphi=w_i\pi+2\pi/3$ with $w_i$ being the independent random number in the range of [-$W$, $W$] ($W\leq2$). Here, $W$ weighs the disorder strength. As shown in Fig. 1d, upon introducing the NNN coupling disorder and increasing its strength, the initially gapless edge state gradually gets opened and midgap modes are generated. The configuration average (100 times) is performed to eliminate the accidental result. To further illustrate the property of this disorder-induced midgap mode, Figs. 1e and 1f display the eigen-spectrum and distribution of midgap mode with $W=1$ ($\varphi=[-\pi/3, 5\pi/3]$). It is clearly shown that the NNN coupling disorders can open a spectral-gap of the original edge state and generate 0D corner states near the zero energy (the slight deviation from the zero-energy is mainly because of the finite size effect. See S1 in the Supporting Information for details). It is worthy to note that the disorder should break the required (crystalline) symmetries for defining the secondary topological index, which is related to the higher-order topological insulator resulting from the filling anomaly [36] (see S2 in the Supporting Information for details). However, we cannot define such a topological invariant in the disordered system. Hence, other definitions of HOTAIs should be found. One appropriate way to define the higher-order topological phase in the disordered system is to adopt the existence of the mid-gap corner state as a working definition [34]. While, one difficulty of this method is that the origin of 0D corner states is often not clear whether they arise from topology or whether they are "accidentally" in the spectrum [37]. Referring to the conventional topological Anderson insulator, where the conductivity is quantized in a certain region of nonzero disorders, we use the disorder-induced quantized fractional corner charge as another convincing evidence to demonstrate the realization of HOTAIs (see S3 in the Supporting Information for details). The green region in Fig. 1d marks the range of disorder strength where the nearly quantized fractional corner charge and mid-gap corner states exist, indicating the realization of the HOTAIs. The physical origin for the disorder-induced corner state can be clarified by considering the competition of two kinds of Dirac masses, where the one is induced by the directional-dependent NNN coupling and the other is caused by the lattice deformation through modulating the intra-cell and inter-cell couplings (see S4 in the Supporting Information for details).

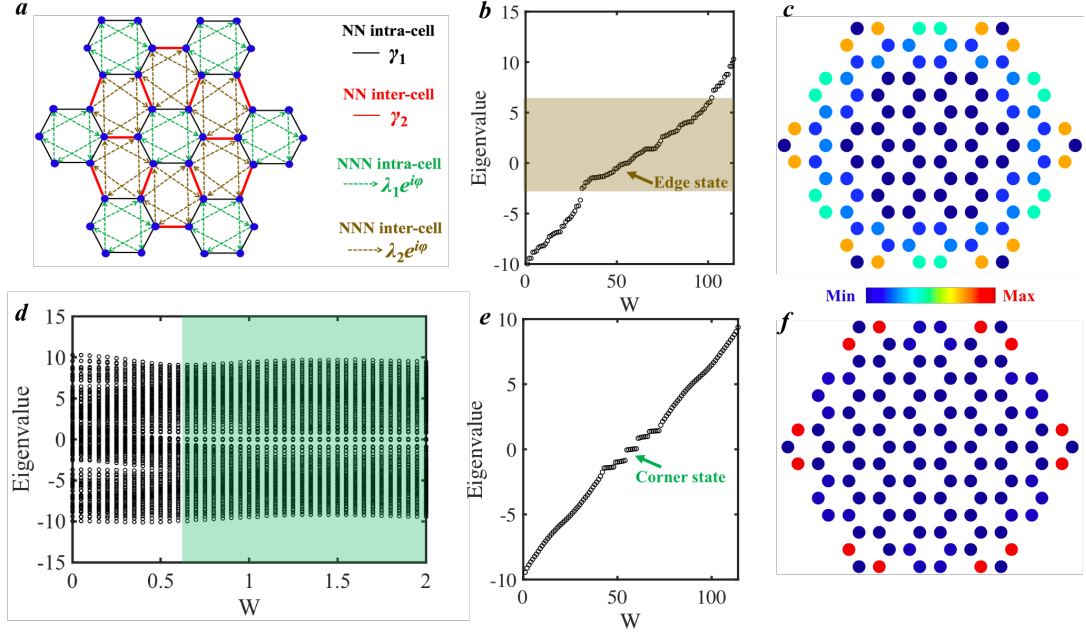

FIG. 1 (a) The schematic diagram of the modified Haldane model. (b) The calculated eigen-spectrum of the open lattice without disorders. The parameters are set as $U=0$, $\gamma_1=1$, $\gamma_2=5$, $\lambda_1=0.4$, $\lambda_2=1.5$, and $\varphi=2\pi/3$. (c) The distribution of zero-energy edge mode without disorders. (d) The eigen-spectrum of open lattice as the function of NNN coupling disorders. The eigen-spectrum (e) and distribution of the zero-energy mode (f) of the open lattice with the disorder strength being $W=1$.

Due to the complex site couplings, which are required to fulfill the HOTAI, constructing the above lattice model is not an easy task no matter using solid materials or classical systems. Recently, based on the similarity between circuit Laplacian and lattice Hamiltonian, simulating topological states with electric circuits has attracted lots of interest [38-47]. Compared with other classical platforms, circuit networks possess remarkable advantages of being versatile and reconfigurable. Consequently, many extremely complex topological states are also fulfilled in circuit networks. In the following, we discuss the experimentally feasible scheme for realizing the HOATI based on electric circuits.

The key to map the modified Haldane model into the circuit network is to construct various types of site-couplings. Four charts in Fig. 2a illustrate the schematic diagram for the construction of four types of site-couplings, respectively. The triangle, which contains three nodes (blue dots) connected by the capacitance $C$ and grounded via a capacitance $C_g$, is considered to form an effective lattice site. The voltages at these three nodes are marked by $V_{i,1}$, $V_{i,2}$, and $V_{i,3}$, which are suitably manipulated to form a pair of pseudospins. To simulate the NN intra-cell (inter-cell) coupling, three inductors with inductances being $L1_{NN}$ ($L2_{NN}$) are used to directly link two triangles labeled by *i* and *j* with $V_{i,a}$ connecting to $V_{j,a}$ (a=1, 2, 3), as shown in the up-left (up-right) chart. As for the realization of NNN intra-cell (inter-cell) couplings, the two triangles are cross-connected via three inductors $L1_{NNN}$ ($L2_{NNN}$), where [$V_{i,1}$, $V_{i,2}$, $V_{i,3}$] is connected to [$V_{j,2}$, $V_{j,3}$, $V_{j,1}$], as presented in the bottom-left (bottom-right) chart. Such a type of cross-connection can introduce a U(1) hopping term accompanied by a geometrical phase [38, 39].

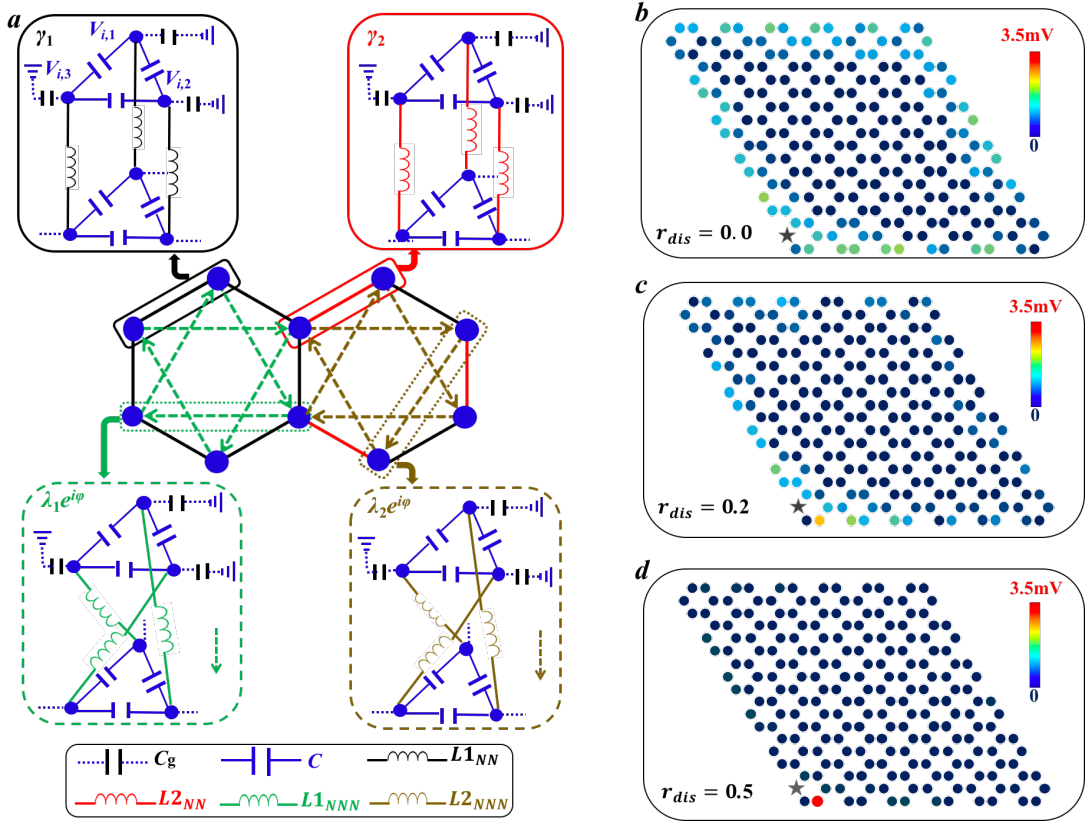

FIG. 2 (a) The schematic diagram for the construction of NN intra-cell, NN inter-cell, NNN intra-cell, and NNN inter-cell couplings in electric circuits. (b)-(d) The voltage distributions at the resonance frequency with the disorder strength being 0, 0.2 and 0.5, respectively.

To present the validity of our designed electric circuit, we derive the eigen-equation of the circuit based on the Kirchhoff's law. By inducing a pair of pseudospins, that are $V_{i,\uparrow}=V_{i,1}+e^{i4\pi/3}V_{i,2}+e^{i2\pi/3}V_{i,3}$ and $V_{i,\downarrow}=V_{i,1}+e^{i2\pi/3}V_{i,2}+e^{i4\pi/3}V_{i,3}$, the two independent equations related to each pseudospin can be written as:

$$EV_{i,\uparrow}=UV_{i,\uparrow}+\frac{1}{L1_{NN}}\sum_{<i,j>}^{intra}V_{i,\uparrow}+\frac{1}{L2_{NN}}\sum_{<i,j>}^{inter}V_{i,\uparrow}+\frac{1}{L1_{NNN}}\sum_{<<i,j>>}^{\varphi>0}e^{i2\pi/3}V_{i,\uparrow}+\frac{1}{L1_{NNN}}\sum_{<<i,j>>}^{\varphi<0}e^{-i2\pi/3}V_{i,\uparrow}$$
$$+\frac{1}{L2_{NNN}}\sum_{<<i,j>>}^{\varphi>0}e^{i2\pi/3}V_{i,\uparrow}+\frac{1}{L2_{NNN}}\sum_{<<i,j>>}^{\varphi<0}e^{-i2\pi/3}V_{i,\uparrow}$$
(2)

$$EV_{i,\downarrow} = UV_{i,\downarrow} + \frac{1}{L1_{NN}}\sum_{<i,j>}^{intra}V_{i,\downarrow} + \frac{1}{L2_{NN}}\sum_{<i,j>}^{inter}V_{i,\downarrow} + \frac{1}{L1_{NNN}}\sum_{<<i,j>>}^{\varphi>0}e^{i2\pi/3}V_{i,\downarrow} + \frac{1}{L1_{NNN}}\sum_{<<i,j>>}^{\varphi<0}e^{-i2\pi/3}V_{i,\downarrow}$$
$$+ \frac{1}{L2_{NNN}}\sum_{<<i,j>>}^{\varphi>0}e^{i2\pi/3}V_{i,\downarrow} + \frac{1}{L2_{NNN}}\sum_{<<i,j>>}^{\varphi<0}e^{-i2\pi/3}V_{i,\downarrow}$$
(3)

where $E=-3\omega^2 C$ and $U=\omega^2 C_g - 2/L1_{NN} - 1/L2_{NN} - 2/L1_{NNN} - 4/L2_{NNN}$. The detailed derivation can be found in the S5 of Supporting Information. For convenience, the grounding capacitance is set as: $C_g = 2C1_{NN} + C2_{NN} + 2C1_{NNN} + 4C2_{NNN}$, where each pair of the LC circuit has the same resonance frequency $1/(L1_{NN}C1_{NN})^{1/2} = 1/(L2_{NN}C2_{NN})^{1/2} = 1/(L1_{NNN}C1_{NNN})^{1/2} = 1/(L2_{NNN}C2_{NNN})^{1/2} = \omega_0$. Eqs. (2) and (3) are the eigenfunction of the modified Haldane model with $U=0$, $\gamma_1 = 1/L1_{NN}$, $\gamma_2 = 1/L2_{NN}$, $\lambda_1 = 1/L1_{NNN}$, $\lambda_2 = 1/L2_{NNN}$ and $\varphi = 2\pi/3$ at $\omega_0$. Hence, it is straightforward to infer that our designed electric circuit can implement the HOTAI by suitably introducing NNN coupling disorders. Here, we chose a simplified scheme to introduce the disorder for the geometric phase of NNN coupling by randomly setting the NNN connection pattern on different circuit nodes that corresponds to randomly set geometric phase as 0, $2\pi/3$ or $4\pi/3$ at different lattice sites. In this case, the disorder strength can be tuned by varying the ratio of nodes with disordered NNN connections. As a general demonstration of disorders induced higher-order topological phases, in the following, we prove that this type of the connection-pattern disorder can bring the system into the higher-order topological phase.

Now, we turn to the electric circuit with open boundaries. Here, a corner-modified (deleting the NNN intra-cell coupling at the corner unit) rhombus lattice is used. It is worthy to note that the grounding of this finite circuit, which is different from the periodic case because of lacking neighbors for edge and corner sites, should be suitably designed to ensure the same on-site energy of each node at the resonance frequency. To validate our scheme, we perform steady-state simulations of the designed open circuit using the LTSpice software. We firstly consider the circuit without disorders. The value of $L1_{NN}$ ($C1_{NN}$), $L2_{NN}$ ($C2_{NN}$), $L1_{NNN}$ ($C1_{NNN}$) and $L2_{NNN}$ ($C2_{NNN}$) are taken as 5uH (0.66nF), 1uH (3.3nF), 10uH (0.33nF) and 3.3uH (1nF). In this case, the resonance frequency of the circuit is 2.77MHz. To obtain the mode response of the circuit, we excite the corner site (marked by the star) with $[V_{i,1}, V_{i,2}, V_{i,3}] = [V_0, V_0 e^{i2\pi/3}, V_0 e^{i4\pi/3}]$ ($V_0$=1V) and calculate the voltage distribution at $f$=2.77MHz, as shown in Fig. 2b. In this case, one of the pseudospins $V_{i,\downarrow}$ is effectively excited. It is clearly shown that dominant voltage signals exist at edge sites and no bulk penetration, manifesting the appearance of edge state belonging to the

AQHP.

Then, we gradually increase the disorder strength of the NNN coupling by increasing the number of nodes with disordered NNN connections ($N_{disorder}$). Figures 2c and 2d present the voltage distribution at 2.77MHz with the disorder strength ($r_{dis}=N_{disorder}/N_{total}$) being 0.2 and 0.5, respectively. Here, ten types of disorder patterns are averaged to eliminate the accidental result. We find that there is still a little voltage signal on the edge, but the voltage on the corner node is significantly increased when the disorder strength of NNN couplings is $r_{dis}=0.2$. And, by further increasing the disorder strength to $r_{dis}=0.5$, the original edge-focused voltage is completely concentrated on the corner, manifesting the edge state is gapped and only the midgap corner state is excited. In this case, we can see that, similar to the tight-binding lattice model, the disorder-induced corner state has been achieved in electric circuits.

To experimentally observe the disorder-induced 0D corner state, we fabricate a series of electric circuits with different disorder strengths. The photograph image of the fabricated sample (without disorders) is shown in Fig. 3a. We can see that the sample contains 6×6 units, which are fabricated on a total of four printed circuit boards (PCBs) with each PCB containing 3×3 units. The sub-PCB is suitably connected to by external circuit elements (white wires). The inset of Fig. 3a presents the enlarged view of the unit cell, where the LC elements are marked in the photo. To ensure the effective excitation of the circuit with a required pseudospin, NI PXIe-8840 Quad-Core Embedded Controller is used to tune the excitation phase and amplitude on three nodes of the corner site. Details of the sample fabrication and voltage measurements are provided in the S6 of Supporting information.

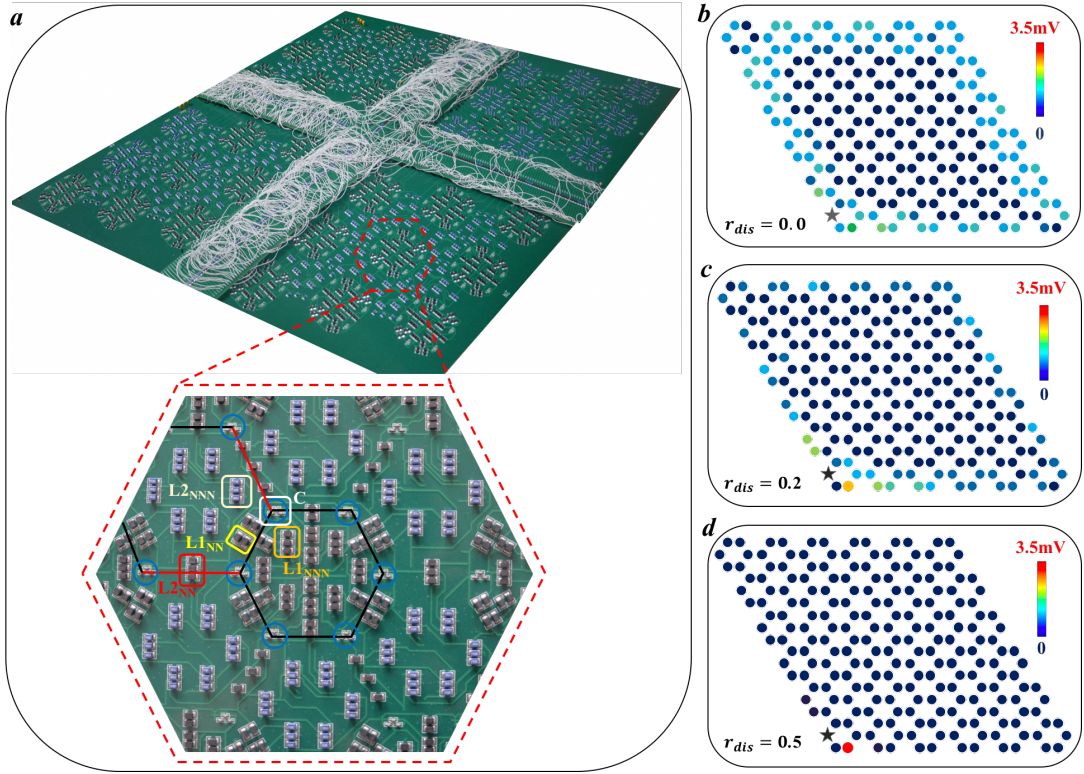

FIG. 3 (a) The photograph of the fabricated electric circuit without disorders. The inset presents the enlarged view of the unit cell. (b)-(d) The measured voltage distribution at 2.77MHz with the disorder strength being 0, 0.2 and 0.5, respectively.

At first, we measure the voltage distribution of the circuit without NNN disorders at 2.77MHz, as shown in Fig. 3b. It is noted that, similar to the simulation results, the dominant voltage signals exist at edge sites, manifesting the system as belonging to the AQHP. Then, we turn to the sample, where the ratio of lattice sites with random NNN connections ($r_{dis}$) equals to 0.2 and 0.5. Figures 3c and 3d display the measured voltage distributions at 2.77MHz with $r_{dis}$=0.2 and 0.5, respectively. Ten types of disorder patterns are averaged to eliminate the accidental result. It is clearly shown that the original edge-dominated voltage signal is gradually concentrated on the corner with increasing the disorder strength. At $r_{dis}$=0.5, the original edge-focused voltage is completely concentrated on the corner, which is consistent with the simulation result in Fig. 2d. In addition, we also measure the voltage signal at corner, edge and bulk sites at different frequencies for both clean and disorder samples. The measured result in the frequency domain also demonstrates the disorder-induced corner mode. See the S7 of Supporting Information for details. In general, both theoretical and experimental results show that corner states can be induced by adding sufficiently strong disorders to the bulk, and the HOTAI can be realized by using the electric circuit.

In conclusion, we have not only demonstrated theoretically that the disorder can induce higher-order topological insulators based on the modified Haldane model, but also realized experimentally the classical analog of HOTAIs carried out in electric circuits. Through the direct circuit simulation and voltage measurement, the 0D corner states induced by the bulk disorder in the 2D system are verified. Such disorder-induced 0D corner states are expected to implement the analog circuit filter that relies on disorders in the bulk. Our work offers a feasible platform to investigate the relationship between disorders and higher-order topological phases, giving rise to the possibility for catching energy at a corner through the import of disorders.

Note: While this manuscript was being prepared, we became aware of three related theoretical works [48-50] that demonstrated the disorder-induced higher-order topological phases.

**Acknowledgements**

This work was supported by the National Key R & D Program of China under Grant No. 2017YFA0303800 and the National Natural Science Foundation of China (No.91850205 and No.61421001).

**References**

[1]. M. Z. Hasan and C. L. Kane, *Rev. Mod. Phys.* **82**, 3045-3067 (2010).

[2]. L. Lu, J. D. Joannopoulos and M. Soljačić, *Nat. Photonics* **8**, 821-829 (2014).

[3]. T. Ozawa, H. M. Price, A. Amo, N. Goldman, M. Hafezi, L. Lu, M. C. Rechtsman, D. Schuster, O. Zilberberg and L. Carusotto, *Rev. Mod. Phys.* **91**, 015006 (2019).

[4]. G. Ma, X. Meng and C. T. Chan, *Nat. Rev. Phys.* **1**, 281-294 (2019).

[5]. G. Harari et al., *Science* **359**, eaar4003 (2018).

[6]. M. A. Bandres et al., *Science* **359**, eaar4005 (2018)

[7]. A. Agarwala and V. B. Shenoy, *Phys. Rev. Lett.* **118**, 236402 (2017).

[8]. W. A. Benalcazar, B. A. Bernevig and T. L. Hughes, *Science* **357**, 61 (2017).

[9]. Z. Song, Z. Fang and C. Fang, *Phys. Rev. Lett.* **119**, 246402 (2017).

[10]. J. Langbehn, Y. Peng, L. Trifunovic, F. von Oppen and P. W. Brouwer, *Phys. Rev. Lett.* **119**, 246401 (2017).

[11]. M. Ezawa, *Phys. Rev. Lett.* **120**, 026801 (2018).

[12]. Q. Zeng, Y. Yang and Y. Xu, *Phys. Rev. B* **101**, 241104(R) (2020).

[13]. F. Schindler et. al, *Nat. Phys.* **14**, 918 (2018).

[14]. F. Schindler et. al, *Sci. Adv.* **4**, eaat0346 (2018).

[15]. M. Serra-Garcia, V. Peri, R. Süsstrunk, O. R. Bilal, T. Larsen, L. G. Villanueva and S. D. Huber, *Nature* **555**, 342 (2018).

[16]. H. Xue, Y. Yang, F. Gao, Y. Chong and B. Zhang, *Nat. Mater.* **18**, 108–112 (2019).

[17]. X. Ni, M. Weiner, A. Alù and A. B. Khanikaev, *Nat. Mater*. **18**, 113–120 (2019).

[18]. C. W. Peterson, W. A. Benalcazar, T. L. Hughes and G. Bahl, *Nature* **555**, 346 (2018).

[19]. S. Mittal, V. V. Orre, G. Zhu, M. A. Gorlach, A. Poddubny and M. Hafezi, *Nat. photonics* **13**, 692-696 (2019).

[20]. W. Zhang et al., *Light: Science & Application* **9**, 109 (2020).

[21]. J. Noh et al., *Nat. Photonics* **12**, 408 (2018).

[22]. S. Imhof et al., *Nat. Phys.* **14**, 925 (2018).

[23]. J. Bao et al., *Phys. Rev. B* **100**, 201406(R) (2019).

[24]. P. W. Anderson, *Phys. Rev.* **109**, 1492–1505 (1958).

[25]. J. Li, R.-L. Chu, J. K. Jain, S.-Q. Shen, *Phys. Rev. Lett.* **102**, 136806 (2009).

[26]. H. Jiang, L. Wang, Q.-F. Sun and X. C. Xie, *Phys. Rev. B* **80**, 165316 (2009).

[27]. C. W. Groth et. al, *Phys. Rev. Lett.* **103**, 196805, (2009).

[28]. H.-M. Guo et. al, *Phys. Rev. Lett.* **105**, 216601, (2010).

[29]. A. Altland et. al, *Phys. Rev. Lett.* **112**, 206602, (2014).

[30]. P. Titum et. al, *Phys. Rev. Lett.* **114**, 056801, (2015).

[31]. E. J. Meier et al, *Science* **362**, 929 (2018).

[32]. F. Zangeneh-Nejad and R. Fleury, *Advanced Materials*, **32**(28), 2001034, (2020).

[33]. S. Stützer et. al, *Nature* **560**, 461 (2018).

[34]. H. Araki, T. Mizoguchi, Y. Hatsugai, *Phys. Rev. B* **99**, 085406 (2019).

[35]. F. D. M. Haldane, *Phys. Rev. Lett.* **61**, 2015 (1988).

[36]. W. A. Benalcazar, T. Li and T. L. Hughes, *Phys. Rev. B*. **99**, 245151 (2019).

[37]. G. V. Miert and C. Ortix, *npj Quantum Materials* **5**, 63 (2020).

[38]. V. Albert, L. I. Glazman, L. Jiang, *Phys. Rev. Lett.* **114**, 173902 (2015).

[39]. J. Ning, C. Owens, A. Sommer, D. Schuster, J. Simon, *Phys. Rev. X* **5**, 021031 (2015).

[40]. N. Olekhno et. al, *Nat. Commun.* **11**, 1436 (2020).

[41]. T. Hofmann, T. Helbig, C. Lee, M. Greiter, R. Thonale, *Phys. Rev. Lett.* **122**, 247702 (2019).

[42]. C. Lee et al, *Communications Physics*, **1**, 39 (2018).

[43]. M. Ezawa, *Phys. Rev. B* **100**, 075423 (2019).

[44]. R. Yu, Y. Zhao, A. P. Schnuder, *National Science Review,* nwaa065, (2020).

[45]. L. Li, C. Lee, J. Gong, *Communications physics* **2**, 135 (2019).

[46]. Y. Wang, H. M. Price, B. Zhang, Y. D. Chong, *Nat. Commun.*, **11**, 2356, (2020).

[47]. W. Zhang et al, Phys. Rev. B, **102**, 100102(R) (2020).

[48]. Franca. S, D. V. Efremov and Fulga, L. C., Phys. Rev. B, 100, 075415, (2019).

[49]. Yang, Y., Li, K., D, L. and Xu, Y. Phys. Rev. B 100, 085408 (2021).

[50]. Li, C., Fu, B., Hu, Z., Li, J. and Shen, S, *Phys. Rev. Lett.* **125**, 166801 (2020).